# Molecular structure retrieval directly from laboratory-frame photoelectron spectra in laser-induced electron diffraction

A. Sanchez[1,*], K. Amini[1,*], S.-J. Wang[2], T. Steinle[1], B. Belsa[1], J. Danek[2], A.T. Le[2,3], X. Liu[1], R. Moshammer[4], T. Pfeifer[4], M. Richter[5], J. Ullrich[4,5], S. Gräfe[6], C.D. Lin[2], J. Biegert[1,7]

[1]ICFO - Institut de Ciencies Fotoniques, The Barcelona Institute of Science and Technology, 08860 Castelldefels (Barcelona), Spain.
[2]Department of Physics, J. R. Macdonald Laboratory, Kansas State University, 66506-2604 Manhattan, KS, USA.
[3]Department of Physics, Missouri University of Science and Technology, Rolla, MO 65409.
[4]Max-Planck-Institut für Kernphysik, Saupfercheckweg 1, 69117, Heidelberg, Germany.
[5]Physikalisch-Technische Bundesanstalt, Bundesallee 100, 38116 Braunschweig, Germany.
[6]Institute of Physical Chemistry and Abbe Center of Photonics, Friedrich-Schiller-Universität Jena, Helmholtzweg 4, 07743 Jena, Germany.
[7]ICREA, Pg. Lluís Companys 23, 08010 Barcelona, Spain.

*Authors contributed equally.

**Abstract**

Ubiquitous to most molecular scattering methods is the challenge to retrieve bond distance and angle from the scattering signals since this requires convergence of pattern matching algorithms or fitting methods. This problem is typically exacerbated when imaging larger molecules or for dynamic systems with little *a priori* knowledge. Here, we employ laser-induced electron diffraction (LIED) which is a powerful means to determine the precise atomic configuration of an isolated gas-phase molecule with picometre spatial and attosecond temporal precision. We introduce a simple molecular retrieval method, which is based only on the identification of critical points in the oscillating molecular interference scattering signal that is extracted directly from the laboratory-frame photoelectron spectrum. The method is compared with a Fourier-based retrieval method, and we show that both methods correctly retrieve the asymmetrically stretched and bent field-dressed configuration of the asymmetric top molecule carbonyl sulfide (OCS), which is confirmed by our quantum-classical calculations.

## I. Introduction

Ultrafast electron diffraction (UED) methods[1–15] are powerful structural probes for gas-phase molecules due to the picometre-scale de Broglie wavelength of the electron, which permits to discern the location of a molecule's atomic constituents. Laser-induced electron diffraction (LIED)[7,9,16] is a strong-field variant of UED, which leverages the laser-controlled recollision of a single electron after tunnelling ionization[1,5,8,11,13,14,16–19] of an attosecond electron wave packet (EWP)[9,14,20], thus uniquely providing picometre and attosecond-to-femtosecond spatial-temporal imaging capabilities. Unlike any other diffraction method, the combination of mid-infrared LIED (MIR-LIED) with a reaction microscope permits the imaging of a single gas-phase molecule with one electron on a single-event basis[9,10,14,21–23]. These unique capabilities present many new opportunities, but they come with the caveat that the strong field itself triggers and modifies the molecular dynamics. In most cases, the influence of the strong field can be described very well with advanced quantum chemistry theory, thus allowing to extract the mechanisms of the processes under investigation despite the field-dressed dynamics. Examples of such treatment are the imaging of molecular bond breaking and deprotonation[10], the photo-dissociation of carbonyl sulfide (OCS)[19], or imaging skeletal deformation due to the Renner Teller effect in carbon disulfide ($CS_2$)[22]. We like to add that presence of the strong field can also be used to our advantage as intra-pulse pump probe method in which the LIED field initiates and probes the dynamics. Variation of the pulse duration gives access to probing different dynamics, e.g. to distinguish between linear or bent stretching dynamics in OCS, or to probe the dynamics of the neutral molecule[22]. Notwithstanding these advances, extraction of the molecular structure from a measured electron interference pattern still requires some sort of pattern matching or minimization procedure, which can lead to molecular parameters with significant variances depending on the exact topography of extremal points in the solution space obtained from measured diffraction images. We note that this is a general challenge for all scattering-based methods, and not unique to LIED. All of these procedures require a degree of background subtraction, de-noising and fitting procedure and, thus, it is highly desirable to minimize any possible variance, which arises from the choice of methodology.

In LIED, the diffraction image is generated by a broadband electron wave packet in the presence of the laser field and typical electron energies range from a few tens to hundreds of eV. This is in contrast to conventional UED in which forward-scattered electrons are measured at small angles (typically a few degrees off the forward direction), but at impact energies of several tens to hundreds of keV. Independent of the method, structural information about the scattering target is encoded into the change of impact momentum, i.e., into the momentum transfer and its range determines the achievable structural resolution. In LIED, this is achieved by measuring low kinetic energy electrons at large scattering angles. Under these conditions, interference of scattering waves from close collisions with all the atoms gives rise to the diffraction images that contain interatomic distances, and the valence electrons have negligible influence on the process. To extract structural information in LIED, we recall that the imaging electron is measured in the laboratory frame by the detector after being accelerated by the linearly polarized laser field and then scattered off the molecular target in the presence of the same laser field. Thus, the laser polarization frame determines the impact direction of the electron, and the momentum transfer that the electron acquires consists of the momentum change due to the scattering target in addition to a momentum shift that is proportional to the laser's vector potential. The additional momentum shift due to the laser can be subtracted from the total momentum, thus making the measured doubly differential scattering cross section (DCS) field free. Based on this laser-driven electron-recollision principle, two methodologies have been developed for LIED, called SD-LIED and FT-LIED, both of which extract structural information from the measured DCS in the laser polarization frame. In SD-LIED, the DCS is extracted for a variety of scattering angles with respect to the laser polarization[5,7,9,18,19] at a fixed scattering electron energy, and is compared to the theoretical DCS calculated from the independent atom model (IAM)[5,24,25]. In this method, bond lengths are obtained using iterative algorithms that find the global minimum in the standard deviation (SD) between the measured and theoretical DCS. Similar SD-based iterative algorithms have also been used in field-free UED and conventional electron diffraction to accurately extract bond lengths within a small forward scattering angle (i.e. < 10°).[12,14,15,26,27] The Fourier-transform LIED (FT-LIED)[14,21,23] method, also called fixed-angle broadband laser-driven electron scattering (FABLES)[8], retrieves the DCS in the backward scattering direction (i.e. along the laser polarization frame, corresponding to 180°) at a range of scattering electron energies. The Fourier transform of the DCS generates the radial distribution of bond lengths. FT-LIED is parameter-free and can directly retrieve the structure of molecules with well-separated bond lengths, however, a broad radial distribution is generated for those with similar bond lengths unless the DCS is accurately measured over a broad range of scattering energies. Even though both SD-LIED and FT-LIED methods have been used successfully for reconstructing the molecular structure of small molecules, the retrieval process is tedious and time consuming. According to the IAM model, the total DCS can be separated into two terms, $\sigma = I_A + I_M$, where $I_A$ is the atomic term, consisting of the incoherent sum of DCS from the single-atom scattering of all the atoms in the molecule, and $I_M$ is the molecular interference term (MIT). The MIT accounts for about 10-15% of the total DCS, and all the information on the molecular structure is contained in the MIT term. Since LIED relies on rescattering process, the total DCS $\sigma$ is already small, but MIT is even smaller. Moreover, this limitation is inherent to all electron diffraction method, with X-ray diffraction being even smaller by a factor of $10^{5-6}$. To advance LIED for probing molecules that are undergoing transitions or retrieving the structure of relatively large molecules, more efficient structural retrieval methods are desirable.

Here, we report an alternative scheme for retrieving the measured molecular structure from LIED experiments by exploiting its two-dimensional data. Firstly, the molecular structure is extracted from the measured MIT term in the laboratory frame to bypass the need to obtain the DCS in the laser frame. Secondly, two-dimensional DCS will be used whenever it is favourable, instead of limiting structural retrieval to one-dimensional DCS data as used in previous SD-LIED and FT-LIED methods. Thirdly, unlike SD-LIED, the molecular structure will be obtained by fitting not the whole MIT term of the DCS, but rather the critical zero crossing points (ZCPs) of the MIT term, making ZCP-LIED superior to SD-LIED as it is simpler and more efficient. We contrast our ZCP-LIED findings against FT-LIED using the same measured data to check if the bond lengths retrieved from a field-dressed carbonyl sulfide (OCS) molecule using both ZCP-LIED and FT-LIED methods are in agreement. We find that ZCP-LIED achieves a higher precision on bond lengths and it overcomes the difficulty of separating bond lengths that are nearly identical in FT-LIED/FABLES, where the radial distributions of bond lengths can overlap significantly. The paper starts with a description of the experiment on the molecular target OCS. Next, we extract the asymmetric top molecule's structure with ZCP-LIED and contrast it against the well-established FT-LIED[9,14,23]/FABLES[8,13] retrieval. Finally, we explain the molecule's bent and asymmetrically stretched configuration with state-of-the-art quantum chemical and dynamical calculations.

## II. Results

The set-up is shown in Fig. 1A, and we refer to Refs[10,20] for further details. Briefly, a gas mixture of 1% OCS seeded in helium was supersonically expanded through a 30-μm aperture, generating a molecular beam with an internal temperature of < 90 K. The molecular beam was collimated using two skimmer stages and subsequently overlapped with a 96 fs (FWHM) 3.2 μm[28,29] laser pulse generated by an optical parametric chirped pulse amplifier (OPCPA) in the interaction region of the reaction microscope. The field strength of the 3.2 μm laser pulse corresponded to a Keldysh parameter of $\gamma \approx 0.3$ and a ponderomotive energy of $U_p = 90$ eV, which led to photo-ionization of electrons with maximal return and back-rescattered electron kinetic energies of $E_{ret,max} = 3.17\ U_p \approx 285$ eV and $E_{resc,max} = 10\ U_p \approx 900$ eV, respectively. The resulting ions and electrons were guided towards separate detectors using homogeneous electric and magnetic extraction fields (19.1 Vcm$^{-1}$ and 10.4 G, respectively). Both the electron and ion detectors consisted of chevron-stacked dual microchannel plates coupled with a quad delay-line anode set-up, enabling the detection of single molecular fragmentation events in full coincidence with 4π acceptance. LIED can be well-described through the framework of laser-driven electron recollision[30–33], as illustrated by the filmstrip inset in Fig. 1a. Here, a molecule is exposed to a strong laser field and an attosecond electron wave packet (EWP) is (i) emitted via tunnel ionization out of the target molecule; (ii) accelerated and returned back to the parent ion by the oscillating electric field of the laser pulse; and (iii) elastically recollides against the target ion where a momentum transfer leaves an imprint of the molecular structure within the EWP's momentum distribution. Figure 1b shows the ion time-of-flight spectrum for all ions. Our analysis in this work will only consider electrons detected in coincidence with the OCS$^+$ ion as part of our LIED imaging process to filter contributions from other processes such as multiple ionization or fragmentation. The two-dimensional momentum map of the transversal and longitudinal momenta (i.e. momenta emitted perpendicular and parallel, respectively, to the laser polarization direction) of electrons detected in coincidence with OCS$^+$ is shown in Fig. 1c. In this map, electrons generated with a longitudinal momentum of > 3 a.u. (< 3 a.u.) correspond to LIED "rescattered" ("direct") electrons detected with a kinetic energy of > 2 U$_p$ ($< 2\ U_p$) that have been returned by the laser field. The schematics shown in Fig. 1c correspond to the data analysis procedure for extracting the scattering differential cross-sections in the laboratory frame (laser polarization frame) for the ZCP-LIED (FT-LIED) method as given by magenta (blue/red) lines and text.

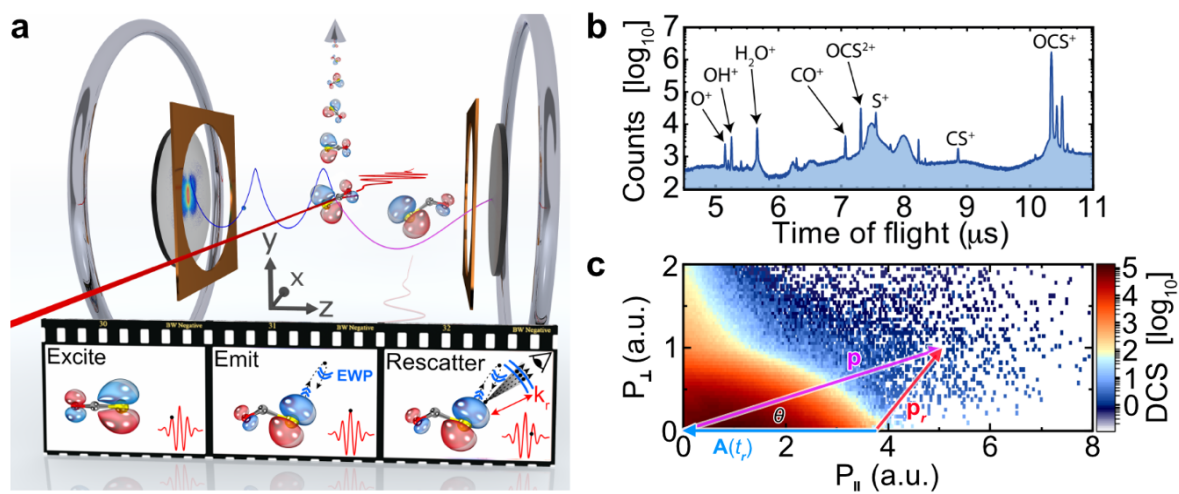

**Fig. 1 LIED experimental set-up and analysis. a** Schematic of laser-induced electron diffraction (LIED) using a reaction microscope coupled with a mid-infrared OPCPA source. The filmstrip inset illustrates the process of imaging the molecular structure with LIED. A molecular jet of OCS molecules is intersected by a 3.2 μm laser pulse that subsequently excites and emits an electron wave packet (EWP; blue lines in filmstrip) before the EWP is returned by the oscillating electric field of the laser pulse and back-rescattered onto the OCS$^+$ target molecule. **b** Measured ion time-of-flight (TOF) spectrum. The OCS$^+$ molecular ion is the most dominant ion TOF peak. **c** Two-dimensional map of transverse ($P_\perp$) and longitudinal ($P_\parallel$) electron momentum distribution (i.e. electron momenta emitted perpendicular and parallel, respectively to the laser polarization direction) detected in coincidence with OCS$^+$ in $\log_{10}$ scale. The momentum of the electron after elastic collision with the ion is depicted by the vector **p** (magenta arrow), while **p$_r$** (red arrow) is the same vector in the laser frame. The two vectors are related by $\mathbf{p} = -\mathbf{A}(t_r) + \mathbf{p_r}$. In the laser frame, the electron undergoes large-angle elastic scattering and emerges in the direction of **p$_r$**. The incident electron is along the polarization axis (horizontal axis) moving toward to the left-hand side. Upon recollision at time $t_r$, it will gain additional momentum from the laser field given by $-\mathbf{A}(t_r)$ (blue arrow) where $\mathbf{A}(t_r)$ is the vector potential of the laser field. It is convenient to think of the laser frame as equivalent to the center-of-mass frame in scattering. The elastic scattering occurs in the laser frame, but the rescattered electrons are measured in the laboratory frame. We use $\theta$ to measure the laboratory-frame angle from the polarization axis. Thus $\theta = 0.0°$ refers to an incident electron that undergoes a 180° backscattering. We consider the laboratory scattering angles of $\theta = 0.0 - 4.0°$ in steps of $\Delta\theta = 0.25°$ to extract the molecular structure. These large-angle (back)scattering events are favorable for probing positions of atoms since they are from collisions at small impact parameters.

Now, we turn to the ZCP-LIED analysis procedure for which we will show that the method allows retrieval of the entire molecular structure directly by identifying zero crossing point (ZCP) positions in the laboratory-frame photoelectron spectrum. First, we average out the local electron counts as a function of scattering momentum, $p$, and scattering angle, $\theta$, (see Fig. 1c). We obtain the laboratory differential cross-section (DCS) for 17 laboratory-frame scattering angles between $0.0 - 4.0°$ in steps of $\Delta\theta = 0.25°$ across the scattering momentum, $p$, range of $4.72 - 7.16$ a.u. (see magenta line with origin $(0,0)$ in Fig. 1c for one exemplary angular distribution that is subsequently shown in Fig. 2a). Including momenta lower than 4.72 a.u. is not considered desirable as the signal begins to rise quickly due to contributions from "direct" electrons and the IAM cannot be applied at small scattering energies, while for higher momenta above 7.16 a.u. the contrast in the modulated signal becomes very small due to an insufficient signal-to-noise ratio. To identify the ZCPs, we first smooth the laboratory DCS with a filter and then fit it with a third-order polynomial function using the least-squares method. The fitted polynomial corresponding to the background atomic $I_A$ signal is then subtracted from the laboratory DCS to obtain the laboratory molecular interference signal, $I_M$. This procedure is repeated for each subsequent angle. The collection of laboratory molecular interference signal $I_M$ at varied θ is shown in Fig. 2a. The amplitude of the interference signal at smaller momenta

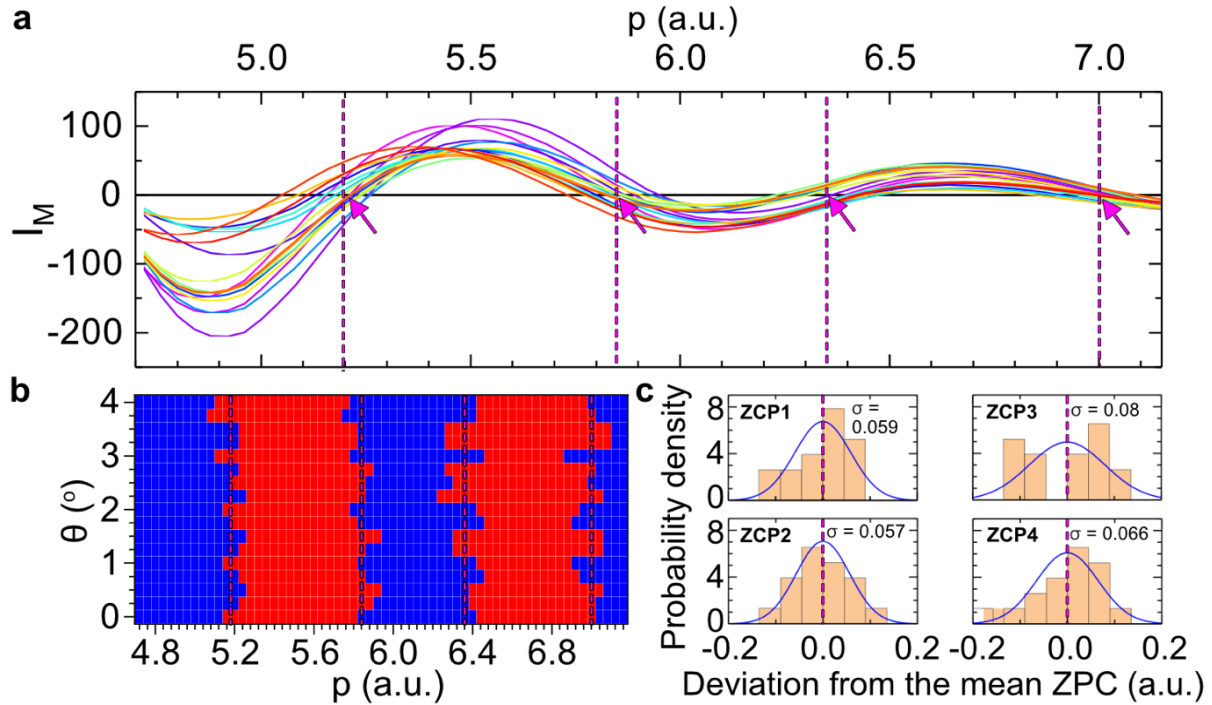

**Fig. 2 Direct retrieval of molecular interference signal from the photoelectron momentum spectrum. a** Molecular interference signal, $I_M$, in the laboratory frame as a function of scattering momentum, $p$, for 17 laboratory scattering angles, between $0.0 - 4.0°$ in steps of $\Delta\theta$=0.25°. Data between $4.72 - 7.16$ a.u. are considered only for statistical reasons and for a valid comparison to the independent atom model (IAM). **b** Two-dimensional map of the negative (blue squares) and positive (red squares) parts of the molecular interference signal as a function of scattering angle and momentum. The magenta dashed vertical lines and arrows in panels (a) and (b) indicate the mean positions of the zero-crossing points (ZCPs) over all scattering angles, which are located at 5.18, 5.84, 6.36 and 7.00

a.u. **c** Frequency of occurrences of finite-width distributions of each averaged experimental ZCP obtained from panel (b). The variances calculated for the four ZCPs are 0.059, 0.057, 0.08 and 0.066 a.u., from the lowest to the highest momenta, respectively.

is larger than that at larger momenta due to the reduced scattering probability with increasing electron scattering momentum. The roots of these molecular interference signals define the ZCPs. The indicated magenta arrows in Fig. 2a and vertical dashed lines in both Figs. 2a and 2b are the mean values of the four ZCPs at four momentum points (5.18, 5.84, 6.36, and 7.00 a.u.) obtained from the measured distributions extracted over the 17 laboratory scattering angles. Fig. 2b provides an alternative view of the two-dimensional (2D) molecular interference signal. We represent positive (negative) values of $I_M$ as red (blue) squares. The boundaries where the $I_M$ changes sign are the locations of the ZCPs. To better visualize the variance, we use the mean value of the ZCP positions as the zero of the momentum axes for all four ZCP distributions. Due to a weaker molecular interference signal at relatively larger scattering momenta where signal-to-noise is lower, the ZCP positions at larger momenta possess a larger error in general. To retrieve the molecular structure, we apply the IAM to obtain the corresponding theoretical ZCPs for a variety of molecular structure parameters. We vary the bond lengths until the square of the difference between the experimental and theoretical ZCP positions is minimal. This optimization process is performed using a genetic algorithm. The optimal internuclear distances found are $R_{CO} = 1.18$ Å, $R_{CS} = 1.70$ Å, and $R_{OS} = 2.74$ Å.

Fig. 3a displays the reconstructed ZCPs generated with the optimal bond lengths at the four momentum points (5.21, 5.82, 6.37, 6.97 a.u.), shown in magenta vertical solid lines, that best match the ZCPs of the experimental data over the 17 scattering angles according to the IAM. These magenta solid lines for the best-fit theoretically calculated ZCPs in Fig. 3a are at momentum positions nearly identical to those measured average ZCPs, shown as magenta vertical dashed lines in Figs. 2a-c (at 5.18, 5.84, 6.36 and 7.00 a.u.). Note that the magenta solid lines in Fig. 3a separate the positive and negative terms of the calculated $I_M$ with no dependence on the scattering angles, which is a general feature predicted by the IAM. Fig. 3b shows the comparison of the experimental molecular interference signals (coloured lines) from 17 scattering angles with the reconstructed ones (black circled line) as a function of the electron momentum in the laboratory frame. We note that the positions of the maxima or minima are fairly independent of scattering angles as well reproduced to a fairly good degree, despite the fact that these positions are not included in the optimization. We comment that the data used in the fitting covers the momentum from 4.7 a.u. to 7.2 a.u. only as in the experimental data analysis procedure described above. The lower limit was chosen at 4.7 a.u. as the IAM fails more as the electron momentum is further reduced. Thus, only four ZCP points are considered in the analysis which are sufficient to accurately retrieve the measured OCS$^+$ molecular structure.

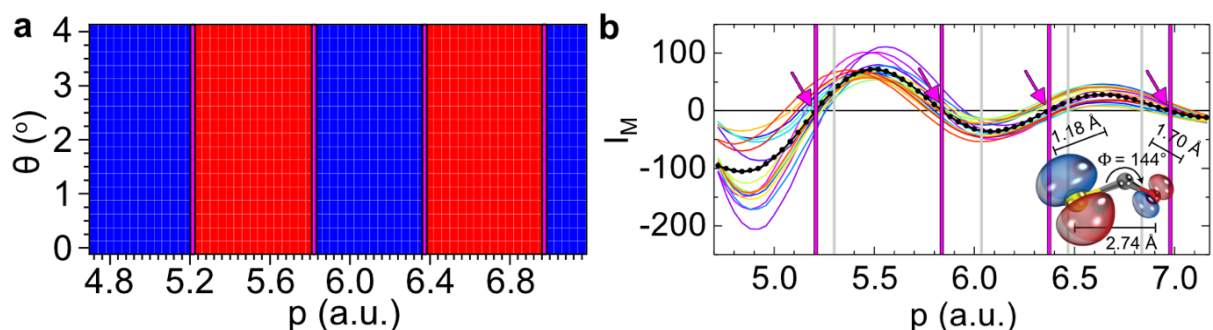

**Fig. 3 ZCP-LIED retrieval of OCS⁺ structure. a** Reconstructed 2D map analogous to that of Fig. 2b. This 2D map was retrieved following the best fit between the experimental data (Fig. 2b) and the corresponding 2D map calculated using the independent atom model (IAM). **b** Molecular interference signal, $I_M$, extracted from experimental data (coloured lines) and the corresponding theoretically calculated distribution (black circled line) that best fits the measured data. A schematic of the reconstructed molecular structure is shown. The zero-crossing points (ZCPs) are indicated by magenta vertical solid lines and arrows. The four reconstructed ZCPs (magenta vertical lines; 5.21, 5.82, 6.37, 6.97 a.u.) from the equilibrium best-fit theoretical data that best match the ZCPs of the experimental data over the 17 scattering angles using the IAM, and correspond to a bent and asymmetrically stretched OCS⁺ structure. The equilibrium neutral ground-state OCS (grey vertical solid lines; 5.30, 6.03, 6.47, 6.84 a.u.) are shown for comparison. Experimental data are shown for $\theta = 0.0 - 4.0°$ in steps of $\Delta\theta = 0.25°$.

Finally, the distribution of experimental ZCP positions for different scattering angles (see Fig. 2c) can be used to estimate the error of the retrieved bond lengths. Since the molecular interference term is evaluated by averaging over a bandwidth of $\Delta p = 0.1$ a.u., Fig. 2c shows the number of angles where the ZCPs falls within 0.1 a.u. From this, we find that around half of the ZCPs fall below or above the magenta dashed line within 0.1 a.u. We can thus estimate the error of each ZCP, and, in turn, calculate the error of the retrieved bond lengths. We extract values of $R_{CO} = 1.18 \pm 0.02$ Å, $R_{CS} = 1.70 \pm 0.02$ Å, and $R_{OS} = 2.74 \pm 0.03$ Å. These distances result in an O-C-S bond angle of $\Phi_{OCS} = 144 \pm 5°$. The theoretical molecular interference signal (black circled line) that best reproduces the measured data (coloured lines), thus, corresponds to a significantly bent and asymmetrically stretched OCS$^+$ structure, as shown in the lower right corner of Fig. 3b, which is drastically different from the equilibrium neutral ground state of OCS. Moreover, the corresponding theoretical ZCPs for equilibrium neutral ground state of OCS appear at four momentum points (grey vertical solid lines; 5.30, 6.03, 6.47, 6.84 a.u.) that are in poor agreement with the measured ZCPs (magenta vertical solid lines; 5.18, 5.84, 6.36, 7.00 a.u.).

Next, we contrast the results from ZCP-LIED against an FT-LIED analysis[21] using the same experimental data as that used in the ZCP-LIED analysis. The experimentally extracted field-free DCS (blue dashed line) is extracted by integrating the electron counts along the return momentum $p_r$ for back-rescattering angles (see red arrow in Fig. 1c) in the laser polarization frame. Fig. 4a shows the extracted DCS with its corresponding error-bars given by the longitudinal detected momentum (red horizontal) and a Poissonian distribution (blue vertical). The oscillations observed in the DCS above 50 eV arise from the coherent molecular interference signal, $I_M$, as a result of the interference between the collision of the EWP and the OCS$^+$ ion. In fact, our measured signal, $I_T$, contains contributions from both $I_M$ and the incoherent sum of atomic scatterings, $I_A$. The latter is independent of the molecular structure and therefore contributes a smooth background signal to our measured $I_T$. This allows the extraction of $I_M$ with a simple subtraction of the empirical background $I_A$ obtained from a linear regression filter to the DCS (black starred line in Fig. 4a). We subsequently enhance the oscillatory modulations in the $I_M$ by contrasting it against the $I_A$ through the molecular contrast factor, MCF $= \frac{I_M}{I_A}$, as shown in Fig. 4b. Here, the typical interference features are present (*i.e.* ZCPs, the local maxima and minima) over a wide momentum transfer range as seen earlier with ZCP-LIED. A simple Fourier transform of the MCF (Fig. 4b) results in Fourier amplitudes which are characteristic of the atomic separations within the measured system, as shown in Fig. 4c. The resulting radial distribution (black dashed line) exhibits three distinct peaks, with the area of uncertainty indicated by the shaded regions corresponding to the detected longitudinal error (red horizontal) and the Poissonian statistical error (blue vertical). The centre positions of these peaks are extracted with a least-squares fit of the measured distribution using the sum of three Gaussian distributions, which correspond to the C-O, C-S and O-S internuclear distances for which we measure values of $R_{CO} = 1.06 \pm 0.06$ Å, $R_{CS} = 1.87 \pm 0.13$ Å, and $R_{OS} = 2.78 \pm 0.10$ Å. These distances result in an O-C-S bond angle of $\Phi_{OCS} = 142 \pm 22°$. We find that these results are in excellent agreement with the results from the ZCP analysis and are summarized in Table 1 together with previously reported structural parameter values for field-free (FF) neutral OCS[34,35] and OCS⁺ cation[36].

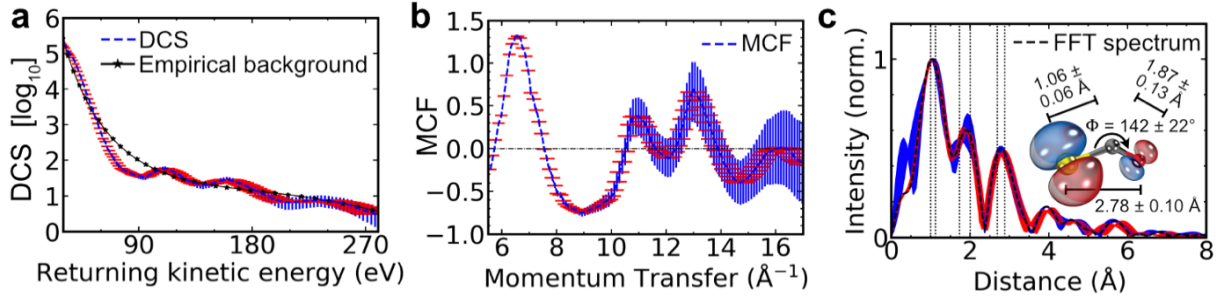

**Fig. 4 FT-LIED retrieval of OCS⁺ structure. a** Differential cross-section (DCS; blue dashed line) as a function of returning kinetic energy corresponding to electrons detected in coincidence with OCS⁺. A linear regression filter (black starred line) is applied to the DCS to extract the incoherent sum of atomic scatterings, $I_A$. The area of uncertainty is given by the longitudinal detected momentum error (red horizontal) and the statistical error based on a Poissonian distribution (blue vertical). **b** The molecular contrast factor (MCF), defined by $I_M/I_A$, as a function of momentum transfer. **c** Radial distribution (black dashed line) obtained by the Fourier transform (FT) of the MCF distribution shown in **b**. The resulting FT spectrum contains three clear peaks corresponding to the C-O, C-S and O-S internuclear distances. The uncertainty in the measured peaks is indicated by the black dotted vertical lines. A schematic of the retrieved asymmetrically stretched and bent OCS⁺ structure is shown in the inset.

|  | $R_{CO}$ (Å) | $R_{CS}$ (Å) | $R_{OS}$ (Å) | $\Phi_{OCS}$ (°) |
|---|---|---|---|---|
| ZCP (this work) | 1.18 ± 0.02 | 1.70 ± 0.02 | 2.74 ± 0.03 | 144 ± 5 |
| FT-LIED (this work) | 1.06 ± 0.06 | 1.87 ± 0.13 | 2.78 ± 0.10 | 142 ± 22 |
| FF OCS ($X^1\Sigma^+$)[34,35] | 1.160 | 1.560 | 2.720 | 180 |
| FF OCS⁺ ($X^2\Pi$)[36] | 1.130 | 1.649 | - | - |

**Table 1 Structural information of OCS and OCS⁺.** The C-O, C-S, and O-S internuclear distances ($R_{CO}$, $R_{CS}$, $R_{OS}$, respectively) with the corresponding O-C-S angle, $\phi_{OCS}$, are presented from our zero-crossing points and FT-LIED retrievals. Corresponding values reported in literature for the equilibrium geometry of neutral field-free (FF) OCS and OCS⁺ in the ground and excited electronic states are presented[32-34].

Finally, we explain the retrieved bent molecular structure and contrast our findings against the linear OCS⁺ structure retrieved at 2.0 μm by Karamatskos *et al.*[19]. Therefore, we perform quantum chemical calculations on the CASSCF/aug-cc-pVTZ level of theory in combination with the classical surface hopping method to describe the dynamics of OCS in the presence of the 3.2 μm laser pulse used in our experiment. In the absence of a field, neutral OCS in its ground electronic state is linear ($C_{\infty v}$)[34,37]. The $1^1\Delta \leftarrow 1^1\Sigma^+$ transition is dipole forbidden due to symmetry. However, field-dressing OCS on the rising edge of the 3.2 μm laser field results in a significantly bent molecule (see Fig. 5c) due to the Renner-Teller effect[22,34,38]. We find in Fig. 5, similar to our recent work in $CS_2$[22] that in the bent geometry, a small population transfer to the $2^1A'$ excited electronic state of the neutral OCS (lower bound of ~3%) occurs. As the ionization potential of this excited state is much lower than the $1^1\Sigma^+$ ground state (~5.2 eV lower), our LIED signal is dominated by neutral OCS in its $2^1A'$ excited state. Consequently, we expect, and measure, a significantly bent and asymmetrically stretched $OCS^+$ structure in a 96 fs (FWHM; 9.0 optical cycles) 3.2 μm field. This is in stark contrast to the recent findings published by Karamatskos et al.[19] who retrieve a linear $OCS^+$ structure. This is accounted for by their much shorter 2.0 μm pulse duration of 38 fs (FWHM; 5.7 optical cycles) being too short in time to invoke the structural deformation of the neutral molecule during the rising edge of their pulse. We chose a longer pulse duration which allows the neutral molecule to undergo significant nuclear deformation on the rising edge of the laser pulse similar to those of $CS_2$ and $C_{60}$.[13,22] We note that our results are in excellent agreement with the work of Sanderson et al. in the Coulomb explosion of $OCS^{n+}$ with a 55 fs (FWHM; 20.6 optical cycles) 0.8 μm laser pulse[39]. These authors assigned the recoil velocities from the ($O^{3+}$, $C^{3+}$, $S^{4+}$) channels to a bent structure ($\Phi_{OCS} = 130 - 140°$) just prior to Coulomb explosion. Recent ab initio calculations at 0.8 μm also revealed that structural deformation in OCS occurs on a time scale longer than 13 fs[40], with wave packet simulations reporting that a linear-to-bent transition occurs on a 50-fs timescale, reaching an OCS bond angle of approximately 135°.[37]

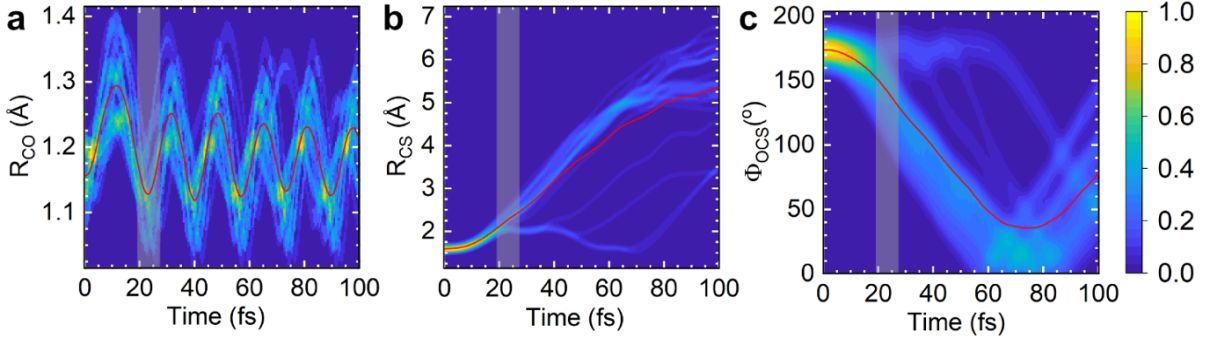

**Fig. 5 Dynamical calculations of field-free neutral OCS. a-c** Mixed quantum-classical nuclear dynamics calculations of neutral field-free OCS that is slightly bent as a function of time after excitation of the neutral molecule, $t$. Dynamical results are shown along the C-O (**a**) and C-S (**b**) internuclear distance coordinates, $R_{CO}$ and $R_{CS}$, respectively, and the O-C-S bond angle (**c**) coordinate, $\Phi_{OCS}$. The average value of over 20 trajectories is shown as a red line. The dynamical calculations best reproduce our measured LIED structure at a time window of $t = 19.5 - 27.0$ fs (shaded grey regions) corresponding to $R_{CO} = 1.13$ Å, $R_{CS} = 2.29$ Å and $\Phi_{OCS} = 141°$.

### III. Discussion and outlook

While the ZCP-LIED method has been demonstrated for the structural retrieval of the OCS molecule only, we believe that the present work offers two important implications which deserve further discussion. The first one is presently limited to LIED experiments only. We have shown the advantage of extracting molecular structure using the full two-dimensional (energy and angle) electron scattering spectra in the laboratory frame, in contrast to existing methods that extract molecular structure in the laser frame. This avoids the tedious procedure of converting the 2D laboratory-frame electron spectra to the laser frame. Using the full 2D scattering spectra instead of 1D spectra improves the statistics of the retrieved results. Moreover, we have demonstrated that ZCP-LIED can retrieve the molecular structure with only the critical crossing points of the molecular interference term (MIT), avoiding the need to use the whole electron scattering spectra directly or the whole MIT. Since molecular bond lengths enter explicitly only in the phase of the interfering terms, by analysing only the ZCPs, the ZCP-LIED method relaxes the limitation of the Born approximation or the independent atom model (IAM) that are at the heart of electron diffraction theory. We further suggest that a combination of ZCP-LIED and FT-LIED methods to retrieve bond lengths from the same scattering data would improve the accuracy (as opposed to FT-LIED only) and avoid false results retrieved based on iterative methods. It should be noted that ZCP-LIED exhibits several advantages over FT-LIED. Firstly, ZCP-LIED retrieves internuclear distances with a higher precision, and it bypasses the challenge of separating internuclear distances that may be hard to distinguish in FT-LIED. Secondly, ZCP-LIED avoids the background polynomial fitting procedure that is used in FT-LIED. Thirdly, ZCP-LIED avoids converting between different frames of reference to retrieve structural information, as is required in FT-LIED and SD-LIED. Finally, ZCP-LIED is capable of structural retrieval at lower signal-to-noise as compared to FT-LIED, and the method is robust in that only a small number of ZCPs are required to accurately retrieve the molecular structure as opposed to requiring the whole MIT or scattering signal for successful retrieval. For example, while we used four ZCPs, in fact only two ZCPs were sufficient to distinguish between the linear (grey vertical lines) and the bent (magenta vertical lines) $OCS^+$ structure (Fig. 3b). Moreover, we have tested the ZCP-LIED method on 10-atom molecules and found that the positions of ZCPs are independent of scattering angles. The accurate, reliable and simplified extraction of molecular structure from scattering data as demonstrated with the ZCP method is essential for the correct interpretation of diffraction imaging-based studies of chemical dynamics and the further evolution of dynamic imaging.

The second important implication from this study is whether the ZCP analysis method can be extended to ultrafast electron diffraction (UED) or even ultrafast X-ray diffraction (UXD). At present, conventional electron and X-ray diffraction methods utilize a monochromatic beam at high energy which can resolve molecular structure with few-picometre precision. For dynamical imaging, high spatial resolution of few-picometre is not needed given that the change in bond length is on the order of tens-to-hundreds of picometres. Would diffraction experiments with a broadband pulse at lower energies be a possible and more advantageous approach for diffraction-based dynamical imaging? Like LIED experiments, if the energy and angle of the scattered electrons can both be measured then two-dimensional differential cross-section data can be determined. This has the advantage that more data are collected in a single experiment. For high-energy beams, the two-dimensional data are redundant since the scattering cross sections depend only on momentum transfer, but similar to what we have demonstrated in this work, such data would provide additional complementary information on the retrieval process. In general, for time-resolved LIED/UED/UXD measurements, we believe that the ZCP analysis will prove to be an important and easy-to-use approach to track the transient geometric structure of a photoexcited molecule by simply tracking the time-resolved changes in the ZCPs of the transient molecular interference signal. Measuring the transient ZCPs will enable a direct retrieval of the transient molecular structure.

In summary, we demonstrate a simple and elegant retrieval of the molecular interference signal directly from the laboratory-frame photoelectron momentum distribution map. We achieve this by identifying the ZCPs of the molecular interference signal which provides a unique fingerprint of molecular structure. Using ZCP-LIED avoids complex retrieval algorithms, semi-classical analyses and *ab initio* calculations to retrieve structural information as is typically used in SD-LIED, UED and UXD. We confirm the ZCP-LIED results with comparison to those obtained using FT-LIED. For both retrievals, we find a significantly bent and asymmetrically stretched $OCS^+$ structure. The observed bent $OCS^+$ structure arises from a linear-to-bent transition assisted by the Renner-Teller effect in the presence of a strong laser field as confirmed by our state-of-the-art quantum-classical dynamical calculations. Avoiding strong deformation of the molecular structure on the rising edge of the 3.2 μm LIED probe pulse as seen in OCS (this work), $CS_2$[22] and $C_{60}$[13] is important in capturing clean snapshots of transient structure in two-pulse pump-probe measurements of photo-induced chemical dynamics, which can be achieved by using 3.2 μm pulses with shorter pulse durations. We expect that ZCP-LIED will prove to be a powerful tool in the retrieval of larger, and more complex, molecular structures as well as transient structures where retrieval algorithms and *ab initio* calculations become increasingly challenging.

**Data availability**

The data and its treatment require dedicated knowledge and experience for a meaningful extraction of information. The data that support the findings of this study are available from the corresponding author upon reasonable request.

**Code availability**

The codes used in this study are available from the corresponding author upon reasonable request.

**Acknowledgements**

J.B. and group acknowledge financial support from the European Research Council for ERC Advanced Grant "TRANSFORMER" (788218), ERC Proof of Concept Grant "miniX" (840010), FET-OPEN "PETACom" (829153), FET-OPEN "OPTOlogic" (899794), Laserlab-Europe (EU-H2020 654148), MINECO for Plan Nacional FIS2017-89536-P; AGAUR for 2017 SGR 1639, MINECO for "Severo Ochoa" (SEV- 2015-0522), Fundació Cellex Barcelona, CERCA Programme / Generalitat de Catalunya, the Polish National Science Center within the project Symfonia, 2016/20/W/ST4/00314, and the Alexander von Humboldt Foundation for the Friedrich Wilhelm Bessel Prize. J.B. and K.A. acknowledge the Polish National Science Center within the project Symfonia, 2016/20/W/ST4/00314, B.B. acknowledges Severo Ochoa" (SEV- 2015-0522), and A.S. acknowledges funding from the Marie Sklodowska-Curie grant agreement No. 641272. S.J.W. and C.D.L are supported in part by Chemical Sciences, Geosciences and Biosciences Division, Office of Basic Energy Sciences, Office of Science, U. S. Department of Energy under Grant No. DE-FG02-86ER13491. M.R. and S.G. highly acknowledges support from the European Research Council (ERC) for the ERC Consolidator Grant QUEM-CHEM (772676).

**Author contributions**

J.B. conceived and supervised the project. A.S., K.A., T.S., X.L. and B.B. performed the measurements. R.M., T.P. and J.U. provided support for the reaction microscope. A.S., K.A. and S.J.W. analyzed the data. M.R. and S.G. performed quantum chemistry calculations. A.T.L. provided cross-section calculations. S. J. W., J.D. and C.D.L. developed the ZCP method. All authors contributed to the data interpretation and writing of the manuscript.

**Competing interests**

The authors declare no competing interests.